\documentclass[conference, a4paper]{IEEEtran}
\IEEEoverridecommandlockouts
\usepackage{cite}
\usepackage{amsmath,amssymb,amsfonts}
\usepackage{algorithmic}
\usepackage{graphicx, xcolor}
\usepackage{textcomp}
\usepackage{comment}
\usepackage{multirow} 
\usepackage{subcaption}

\def\BibTeX{{\rm B\kern-.05em{\sc i\kern-.025em b}\kern-.08em
    T\kern-.1667em\lower.7ex\hbox{E}\kern-.125emX}}
\begin{document}

\title{Image Transformation Network \\ for Privacy-Preserving Deep Neural Networks \\ and Its Security Evaluation}


\author{\IEEEauthorblockN{1\textsuperscript{st} Hiroki Ito}
\IEEEauthorblockA{\textit{Department of Computer Science} \\
\textit{Tokyo Metropolitan University}\\
Tokyo, Japan \\
ito-hiroki2@ed.tmu.ac.jp}
\and
\IEEEauthorblockN{2\textsuperscript{nd} Yuma Kinoshita}
\IEEEauthorblockA{\textit{Department of Computer Science} \\
\textit{Tokyo Metropolitan University}\\
Tokyo, Japan \\
ykinoshita@tmu.ac.jp}
\and
\IEEEauthorblockN{3\textsuperscript{rd} Hitoshi Kiya}
\IEEEauthorblockA{\textit{Department of Computer Science} \\
\textit{Tokyo Metropolitan University}\\
Tokyo, Japan \\
kiya@tmu.ac.jp}
}

\maketitle

\begin{abstract}
	We propose a transformation network for generating visually-protected images for privacy-preserving DNNs.
The proposed transformation network is trained by using a plain image dataset so that plain images are transformed into visually protected ones. 
Conventional perceptual encryption methods have a weak visual-protection performance and some accuracy degradation in image classification. 
In contrast, the proposed network enables us not only to strongly protect visual information but also to maintain the image classification accuracy that using plain images achieves. 
In an image classification experiment, the proposed network is demonstrated to strongly protect visual information on plain images without any performance degradation under the use of CIFAR datasets.
In addition, it is shown that the visually protected images are robust against a DNN-based attack, called inverse transformation network attack (ITN-Attack) in an experiment.
\end{abstract}

\begin{IEEEkeywords}
deep neural network, privacy preserving, visual protection
\end{IEEEkeywords}

\section{Introduction}
	The spread use of deep neural networks (DNNs) has greatly contributed to solving complex tasks for many applications \cite{dl_1, dl_2}, including privacy-sensitive/security-critical ones such as facial recognition and medical image analysis.
Recently, it has been very popular for data owners to utilize cloud servers to compute and process a large amount of data instead of using local servers.
However, there are risks of data leakage in the cloud environment \cite{cloud}.
Because application users (i.e.\ clients) want to avoid the risks, privacy-preserving DNNs have become an urgent challenge.
In this paper, we focus on protecting visual information on images before uploading them to cloud environments.

Perceptual encryption generates images that can be directly applied to various image processing algorithms, but information theory-based encryption (like RSA and AES) generates a ciphertext.
In the past years, various perceptual encryption methods have already been proposed \cite{vp_1, vp_2, vp_3, vp_4, vp_ml_1, vp_ml_3, vp_ml_4, vp_ml_6, vp_ml_65, vp_ml_7, seo_pre, tanaka, pixel-based, pixel-based2, pixel2, gan-protect}.
In these methods, there are only three methods for privacy-preserving DNNs: Tanaka's method \cite{tanaka}, a pixel-based encryption method \cite{pixel-based, pixel-based2}, and a generative adversarial network (GAN)-based method using an image transformation network \cite{gan-protect}.
However, the use of the methods degrades the performance of DNNs, compared with the use of plain images.

For such reasons, in this paper, we propose a transformation network for generating visually-protected images for privacy-preserving DNNs.
The proposed network transforms a plain image into a visually-protected one.
The proposed network is trained so that generated images reduce the loss value of a classification model.

Experiments using the CIFAR-10 and 100 datasets \cite{cifer10} show that the proposed network enables us not only to protect visual information on plain images but also to maintain the performance of DNNs.
Furthermore, we demonstrate that visual information on plain images can not be restored from visually-protected images by a DNN-based attack, called inverse transformation network attack (ITN-Attack).

\section{Proposed Transformation Network}
	\subsection{Overview}
\begin{figure}[t!]
	\centering
 	\centerline{\includegraphics[width=8.5cm] {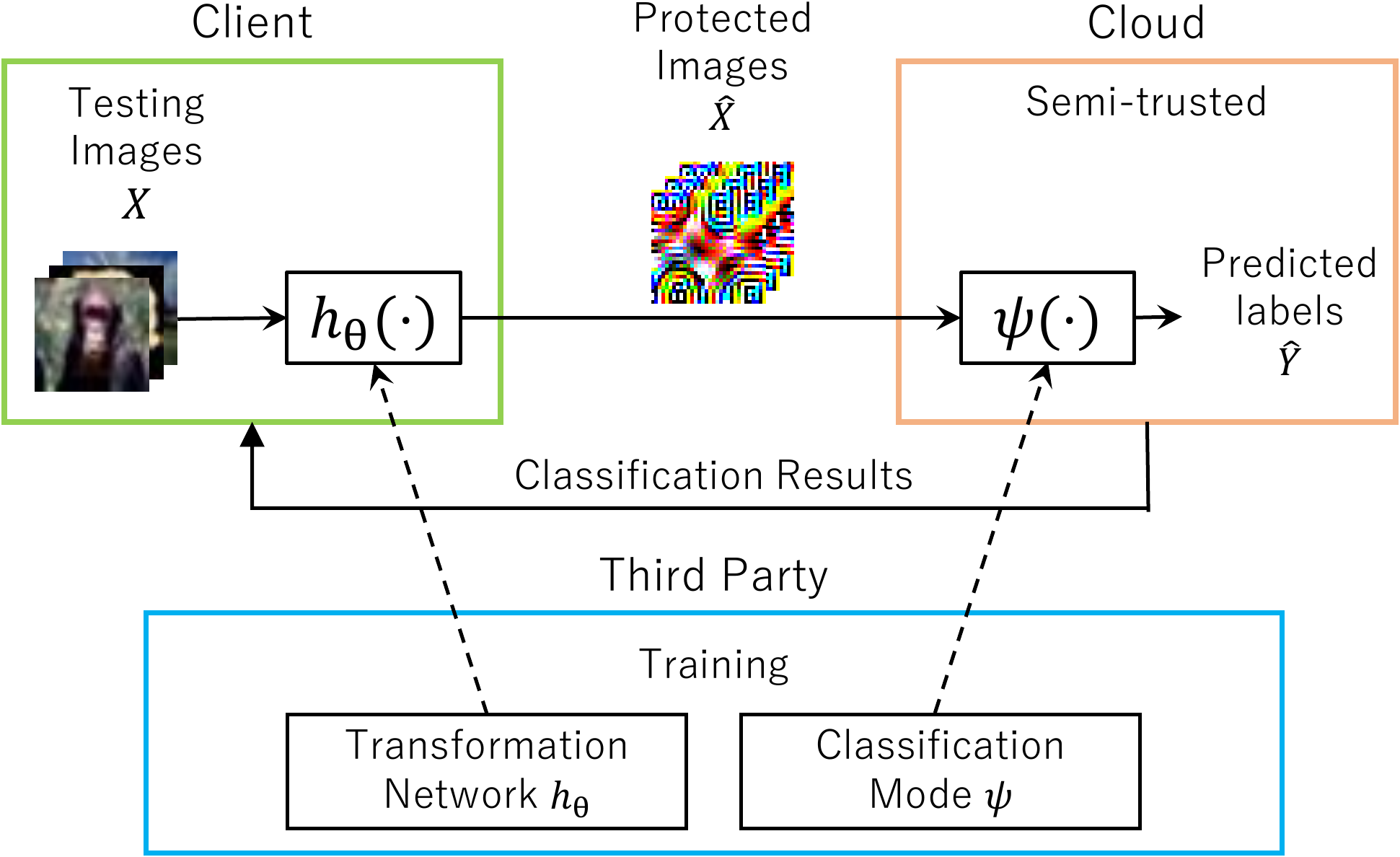}}
  	\caption{Framework of proposed scheme}
	\label{fig:proposed_overview}
\end{figure}

Figure \ref{fig:proposed_overview} illustrates the framework that we assume in this paper.
In this framework, transformation network $h_\theta$ is public to clients, and classification model $\psi$ is available on a cloud server.
The client sends visually-protected images generated by using $h_\theta$ to the cloud server.
The cloud server classifies the images by using model $\psi$ and returns the results to the client.
In this framework, the cloud server has no visual information on plain images, so visual information is protected even if the cloud server is not trusted.

\subsection{Training Transformation Network}
\begin{figure}[t!]
	\centering
 	\centerline{\includegraphics[width=8.5cm] {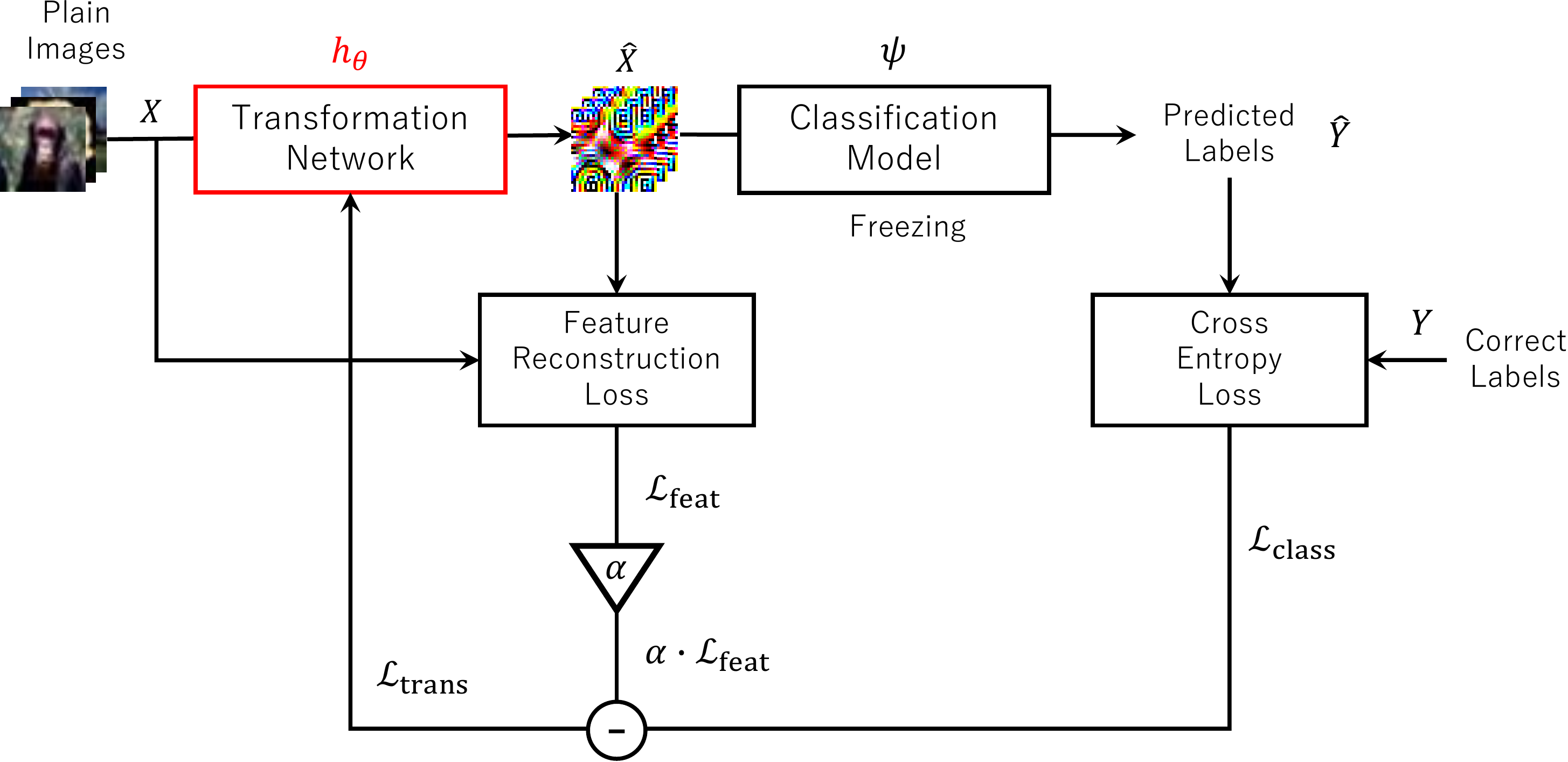}}
  	\caption{Training process of transformation network $h_\theta$}
	\label{fig:conversion_train}
\end{figure}

The training procedure of the proposed transformation network is illustrated in Fig.\ \ref{fig:conversion_train}, 
where $X = \{ x_1, x_2, \ldots, x_m \}$ is an input plain image set, 
$\hat{X} = \{ \hat{x}_1, \hat{x}_2, \ldots, \hat{x}_m\}$ is an output image set 
from the transformation network, i.e., $\hat{x}_i = h_\theta(x_i)$, 
$Y = \{ y_1, y_2, \ldots, y_m \}$ is a one-hot encoded target label set,
and $\hat{Y} = \{ \hat{y}_1, \hat{y}_2, \ldots, \hat{y}_m\}$ is an output label set 
from a classification network, i.e., $\hat{y}_i = \psi(\hat{x}_i)$.
One-hot encoded label, $y_i = (y_i(1), y_i(2), \ldots, y_i(c))$ and output label $\hat{y}_i = (\hat{y}_i(1), \hat{y}_i(2), \ldots, \hat{y}_i(c))$ meet 
\begin{gather}
    y_i(j) \in \{ 0, 1\}, \text{and} \sum_{j=1}^{c} y_i(j) = 1.
\end{gather}
and
\begin{gather}
    0 \leq \hat{y}_i(j) \leq 1, \text{and} \sum_{j=1}^{c} \hat{y}_i(j) = 1,
\end{gather}
respectively, where $c$ is the number of classes.
The proposed network converts images to visually protected ones.
Network $h_\theta$ is trained so that generated images reduce the loss value of model $\psi$.

To train network $h_\theta$ with parameter $\theta$ 
by using a plain input image $x_i$ and its one-hot encoded target label $y_i$, 
loss function $\mathcal{L}_{\mathrm{trans}}$ is minimized as
\begin{equation}
	\underset{\theta}{\mathrm{minimize}} \quad \frac{1}{m} \sum_{i=1}^{m} \mathcal{L}_{\mathrm{trans}}(x_i, h_\theta(x_i), y_i),
\end{equation}
with
\begin{equation}
\label{eq:trans-loss}
	\mathcal{L}_{\mathrm{trans}}(x_i, \hat{x}_i, y_i) = \mathcal{L}_{\mathrm{class}}(\hat{x}_i, y_i) 
	- \alpha \cdot \mathcal{L}_{\mathrm{feat}}(x_i, \hat{x}_i),
\end{equation}
where $\mathcal{L}_{\mathrm{class}}$ denotes a classification loss function,
which is used to classify visually protected images correctly, 
$\mathcal{L}_{\mathrm{feat}}$ is a feature reconstruction loss function to be used for visually protecting input images, and $\alpha \in \mathbb{R}$ is a weight of $\mathcal{L}_{\mathrm{feat}}$.

In this paper,
$\mathcal{L}_{\mathrm{class}}$ is given by the cross-entropy loss.
Therefore, $\mathcal{L}_{\mathrm{class}}$ is calculated by using $\hat{y}_i(j)$ as
\begin{equation}
    \mathcal{L}_{\mathrm{class}}(\hat{x}_i, y_i) =
        -\sum_{j=1}^{c} y_i(j) \log \hat{y}_i(j).
\end{equation}
$\mathcal{L}_{\mathrm{feat}}$ is also given by
\begin{equation}
    \mathcal{L}_{\mathrm{feat}}(x_i, \hat{x}_i) =
        \frac{1}{C_k H_k W_k} \| \phi_k(\hat{x}_i) - \phi_k(x_i) \| _2^2,
\end{equation}
where $\phi_k(x)$ is a feature map
with a size of $C_k \times H_k \times W_k$
obtained by the $k$-th layer of a network when image $x$ is fed \cite{perceptual}.

\subsection{Robustness against DNN-based Attacks}
One of the state-of-the-art attacks is a GAN (generative adversarial network)-based one \cite{gan-attack}. 
The GAN-based attack may enable us to estimate visual information on plain images from visually-protected images without a correct pair set of plain images and protected images in general. 
Although, in our scheme (see Fig.\ \ref{fig:proposed_overview}), attackers can easily prepare a correct set because $h_\theta$ is open to the public. 
Therefore, attackers can create an inverse transformation network more efficiently by using a correct pair set for estimating visual information on plain images. 
In this paper, we tried to train an inverse transformation network by using a correct pair set. 
We call it an inverse transformation network attack (ITN-Attack). 
Even when ITN-Attack is applied to protected images generated by using $h_\theta$, the protected ones will be shown to be robust enough against ITN-Attack in an experiment.

\section{Simulations}
	We evaluated the proposed transformation network in terms of classification accuracy and visual protection performance.
Robustness against ITN-Attack was also evaluated.

\subsection{Evaluating Transformation Network Performance}
We used U-Net \cite{u-net} and ResNet-20 \cite{resnet} as transformation network $h_\theta$ and classification model $\psi$, respectively.
Also, we used the CIFAR-10 and 100 datasets \cite{cifer10}.
Each dataset consists of a training set with 50,000 images and a test set with 10,000.
In experiments using CIFAR-10, we utilized 45,000 images in the training set to train both $\psi$ and $h_\theta$, and the other 5,000 images were used as validation data.
In contrast, we utilized 47,500 images for training both networks in experiments using CIFAR-100, and the other 2,500 images were used as validation data.
The test set of each dataset was also utilized for evaluating the performance of the proposed method.
In addition, standard data-augmentation methods, i.e., random crop and horizontal flip, were performed in the training.

All networks were trained for 200 epochs,
by using the stochastic gradient descent (SGD) with
a weight decay of 0.0005 and a momentum of 0.9.
The learning rate was initially set to 0.1
and it was multiplied by 0.2 at 60, 120, and 160 epochs.
The batch size was 128.
After the training,
we selected the network
that provided the lowest loss value
under the use of the validation set.

\begin{figure*}[t!]
	\centering
	\begin{subfigure}[t]{\hsize}
		\centering
		\begin{subfigure}[t]{0.09\hsize}
			\centering
			\includegraphics[width=1.5cm]{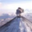}
		\end{subfigure}
		\begin{subfigure}[t]{0.09\hsize}
			\centering
			\includegraphics[width=1.5cm]{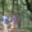}
		\end{subfigure}
		\begin{subfigure}[t]{0.09\hsize}
			\centering
			\includegraphics[width=1.5cm]{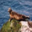}
		\end{subfigure}
		\begin{subfigure}[t]{0.09\hsize}
			\centering
			\includegraphics[width=1.5cm]{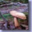}
		\end{subfigure}
		\begin{subfigure}[t]{0.09\hsize}
			\centering
			\includegraphics[width=1.5cm]{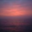}
		\end{subfigure}
		\begin{subfigure}[t]{0.09\hsize}
			\centering
			\includegraphics[width=1.5cm]{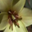}
		\end{subfigure}
		\begin{subfigure}[t]{0.09\hsize}
			\centering
			\includegraphics[width=1.5cm]{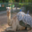}
		\end{subfigure}
		\begin{subfigure}[t]{0.09\hsize}
			\centering
			\includegraphics[width=1.5cm]{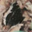}
		\end{subfigure}
		\begin{subfigure}[t]{0.09\hsize}
			\centering
			\includegraphics[width=1.5cm]{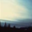}
		\end{subfigure}
		\begin{subfigure}[t]{0.09\hsize}
			\centering
			\includegraphics[width=1.5cm]{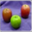}
		\end{subfigure} 
		\caption{Plain}
	\end{subfigure}
	\\ \vspace{1mm}
	\begin{subfigure}[t]{\hsize}
		\centering
		\begin{subfigure}[t]{0.09\hsize}
			\centering
			\includegraphics[width=1.5cm]{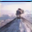}
		\end{subfigure}
		\begin{subfigure}[t]{0.09\hsize}
			\centering
			\includegraphics[width=1.5cm]{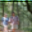}
		\end{subfigure}
		\begin{subfigure}[t]{0.09\hsize}
			\centering
			\includegraphics[width=1.5cm]{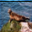}
		\end{subfigure}
		\begin{subfigure}[t]{0.09\hsize}
			\centering
			\includegraphics[width=1.5cm]{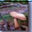}
		\end{subfigure}
		\begin{subfigure}[t]{0.09\hsize}
			\centering
			\includegraphics[width=1.5cm]{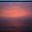}
		\end{subfigure}
		\begin{subfigure}[t]{0.09\hsize}
			\centering
			\includegraphics[width=1.5cm]{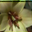}
		\end{subfigure}
		\begin{subfigure}[t]{0.09\hsize}
			\centering
			\includegraphics[width=1.5cm]{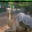}
		\end{subfigure}
		\begin{subfigure}[t]{0.09\hsize}
			\centering
			\includegraphics[width=1.5cm]{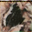}
		\end{subfigure}
		\begin{subfigure}[t]{0.09\hsize}
			\centering
			\includegraphics[width=1.5cm]{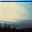}
		\end{subfigure}
		\begin{subfigure}[t]{0.09\hsize}
			\centering
			\includegraphics[width=1.5cm]{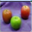}
		\end{subfigure} 
		\caption{Protected ($\alpha = 0$)}
	\end{subfigure}
	\\ \vspace{1mm}
	\begin{subfigure}[t]{\hsize}
		\centering
		\begin{subfigure}[t]{0.09\hsize}
			\centering
			\includegraphics[width=1.5cm]{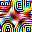}
		\end{subfigure}
		\begin{subfigure}[t]{0.09\hsize}
			\centering
			\includegraphics[width=1.5cm]{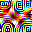}
		\end{subfigure}
		\begin{subfigure}[t]{0.09\hsize}
			\centering
			\includegraphics[width=1.5cm]{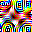}
		\end{subfigure}
		\begin{subfigure}[t]{0.09\hsize}
			\centering
			\includegraphics[width=1.5cm]{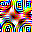}
		\end{subfigure}
		\begin{subfigure}[t]{0.09\hsize}
			\centering
			\includegraphics[width=1.5cm]{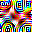}
		\end{subfigure}
		\begin{subfigure}[t]{0.09\hsize}
			\centering
			\includegraphics[width=1.5cm]{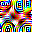}
		\end{subfigure}
		\begin{subfigure}[t]{0.09\hsize}
			\centering
			\includegraphics[width=1.5cm]{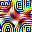}
		\end{subfigure}
		\begin{subfigure}[t]{0.09\hsize}
			\centering
			\includegraphics[width=1.5cm]{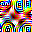}
		\end{subfigure}
		\begin{subfigure}[t]{0.09\hsize}
			\centering
			\includegraphics[width=1.5cm]{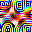}
		\end{subfigure}
		\begin{subfigure}[t]{0.09\hsize}
			\centering
			\includegraphics[width=1.5cm]{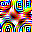}
		\end{subfigure} 
		\caption{Protected ($\alpha = 0.005$)}
	\end{subfigure}
	\\ \vspace{1mm}
	\begin{subfigure}[t]{\hsize}
		\centering
		\begin{subfigure}[t]{0.09\hsize}
			\centering
			\includegraphics[width=1.5cm]{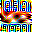}
		\end{subfigure}
		\begin{subfigure}[t]{0.09\hsize}
			\centering
			\includegraphics[width=1.5cm]{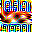}
		\end{subfigure}
		\begin{subfigure}[t]{0.09\hsize}
			\centering
			\includegraphics[width=1.5cm]{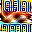}
		\end{subfigure}
		\begin{subfigure}[t]{0.09\hsize}
			\centering
			\includegraphics[width=1.5cm]{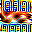}
		\end{subfigure}
		\begin{subfigure}[t]{0.09\hsize}
			\centering
			\includegraphics[width=1.5cm]{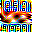}
		\end{subfigure}
		\begin{subfigure}[t]{0.09\hsize}
			\centering
			\includegraphics[width=1.5cm]{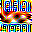}
		\end{subfigure}
		\begin{subfigure}[t]{0.09\hsize}
			\centering
			\includegraphics[width=1.5cm]{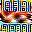}
		\end{subfigure}
		\begin{subfigure}[t]{0.09\hsize}
			\centering
			\includegraphics[width=1.5cm]{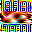}
		\end{subfigure}
		\begin{subfigure}[t]{0.09\hsize}
			\centering
			\includegraphics[width=1.5cm]{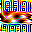}
		\end{subfigure}
		\begin{subfigure}[t]{0.09\hsize}
			\centering
			\includegraphics[width=1.5cm]{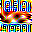}
		\end{subfigure} 
		\caption{Protected ($\alpha = 0.01$)}
	\end{subfigure}
	\\
  	\caption{Visually protected images generated
  	    by proposed transformation network trained with ResNet-20}
	\label{fig:sample_protected}
\end{figure*}

Figure \ref{fig:sample_protected} shows
an example of visually protected images generated from ten test images in CIFAR-100,
by using $h_\theta$,
where the top row shows plain images
and the second top row to bottom row shows
images generated with the parameters 
$\alpha=0$, 0.005, and 0.01 in Eq.\ (\ref{eq:trans-loss}).

From the figure, the generated images had almost no visual information on the plain images when $\alpha \geq 0.005$.
Also, in the case of $\alpha = 0$, the generated images were not visually protected, since a loss for visually protecting input images ($\mathcal{L}_{\mathrm{feat}}$) did not work.
Also, all protected images have a similar pattern.
Thus, the protected images have almost no visual information on the plain images in addition to the high classification accuracy.

Table \ref{tab:ab_accuracy} shows the classification accuracy when the generated images were protected.
From the table, when $\alpha \leq 0.01$, the proposed network provided higher classification accuracy than conventional methods.
The reason that the accuracy improved is that the proposed network increases the total number of parameters due to the use of the transformation network.
\begin{table}[t!]
	\centering
	\caption{Classification accuracy (\%)}
	\begin{tabular}{c|l|c|c} \hline \hline
		\multicolumn{2}{c|}{Method} & CIFAR-10 & CIFAR-100 \\ \hline
		\multirow{4}{*}{Proposed} & $\alpha=0.005$  & \boldmath $91.72$ & \boldmath $70.78$ \\
		& $\alpha=0.01$    & \boldmath $91.41$ & \boldmath $70.08$ \\
		& $\alpha=0.05$    & 89.63                  & 42.91 \\         
		& $\alpha=0.1$     & 39.92                 & 1.00 \\ \hline 
		\multicolumn{2}{c|}{Plain image} & 91.23 & 67.9 \\
             \multicolumn{2}{c|}{Tanaka \cite{tanaka}}  & 85.18 & 60.08 \\
             \multicolumn{2}{c|}{Pixel-based \cite{pixel-based, pixel-based2}} & 90.99 & 60.50 \\ \hline \hline
	\end{tabular}
	\label{tab:ab_accuracy}
\end{table}

\subsection{Evaluating Robustness against ITN-Attack}
\begin{figure}[t!]
	\centering
	\begin{subfigure}[t]{0.95\hsize}
	\centering
	\begin{subfigure}[t]{0.23\hsize}
		\centering
		\includegraphics[width=1.5cm]{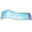}
	\end{subfigure}
	\hfill
	\begin{subfigure}[t]{0.23\hsize}
		\centering
		\includegraphics[width=1.5cm]{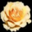}
	\end{subfigure}
	\hfill
	\begin{subfigure}[t]{0.23\hsize}
		\centering
		\includegraphics[width=1.5cm]{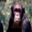}
	\end{subfigure} 
	\hfill
	\begin{subfigure}[t]{0.23\hsize}
		\centering
		\includegraphics[width=1.5cm]{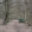}
	\end{subfigure}
	\caption{Plain}
	\end{subfigure}
	\\
	\begin{subfigure}[t]{0.95\hsize}
	\centering
	\begin{subfigure}[t]{0.23\hsize}
		\centering
		\includegraphics[width=1.5cm]{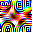}
		\caption*{$\rm{PSNR}=3.86$}
	\end{subfigure}
	\hfill
	\begin{subfigure}[t]{0.23\hsize}
		\centering
		\includegraphics[width=1.5cm]{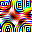}
		\caption*{$\rm{PSNR}=4.96$}
	\end{subfigure}
	\hfill
	\begin{subfigure}[t]{0.23\hsize}
		\centering
		\includegraphics[width=1.5cm]{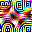}
		\caption*{$\rm{PSNR}=5.91$}
	\end{subfigure}
	\hfill
	\begin{subfigure}[t]{0.23\hsize}
		\centering
		\includegraphics[width=1.5cm]{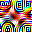}
		\caption*{$\rm{PSNR}=7.11$}
	\end{subfigure}
	\caption{Protected}
	\end{subfigure}
	\\
	\begin{subfigure}[t]{0.95\hsize}
	\centering
	\begin{subfigure}[t]{0.23\hsize}
		\centering
		\includegraphics[width=1.5cm]{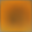}
		\caption*{$\rm{PSNR}=4.58$}
	\end{subfigure}
	\hfill
	\begin{subfigure}[t]{0.23\hsize}
		\centering
		\includegraphics[width=1.5cm]{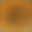}
		\caption*{$\rm{PSNR}=8.23$}
	\end{subfigure}
	\hfill
	\begin{subfigure}[t]{0.23\hsize}
		\centering
		\includegraphics[width=1.5cm]{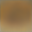}
		\caption*{$\rm{PSNR}=10.7$}
	\end{subfigure}
	\hfill
	\begin{subfigure}[t]{0.23\hsize}
		\centering
		\includegraphics[width=1.5cm]{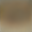}
		\caption*{$\rm{PSNR}=21.5$}
	\end{subfigure}
	\caption{Estimated}
	\end{subfigure}
	\caption{Estimated images by inverse transformation network with $h_\theta$ trained with CIFAR-100 and $\alpha = 0.005$}
	\label{fig:enc-stl10}
\end{figure}

In Fig. \ref{fig:enc-stl10}, images estimated by using the inverse transformation model are illustrated together with the corresponding plain images and the visually protected ones.
The inverse transformation model was trained by using $h_\theta$ trained with $\alpha = 0.005$.
To evaluate the error of the estimation, peak signal-to-noise ratio (PSNR) between estimated images and the plain images were also calculated (see the bottom of each image).
From Fig.\ \ref{fig:enc-stl10}, the estimated images had almost no visual information on the plain images and most estimated images had low PSNR values.

\begin{figure}[t!]
	\centering
 	\centerline{\includegraphics[width=8.5cm] {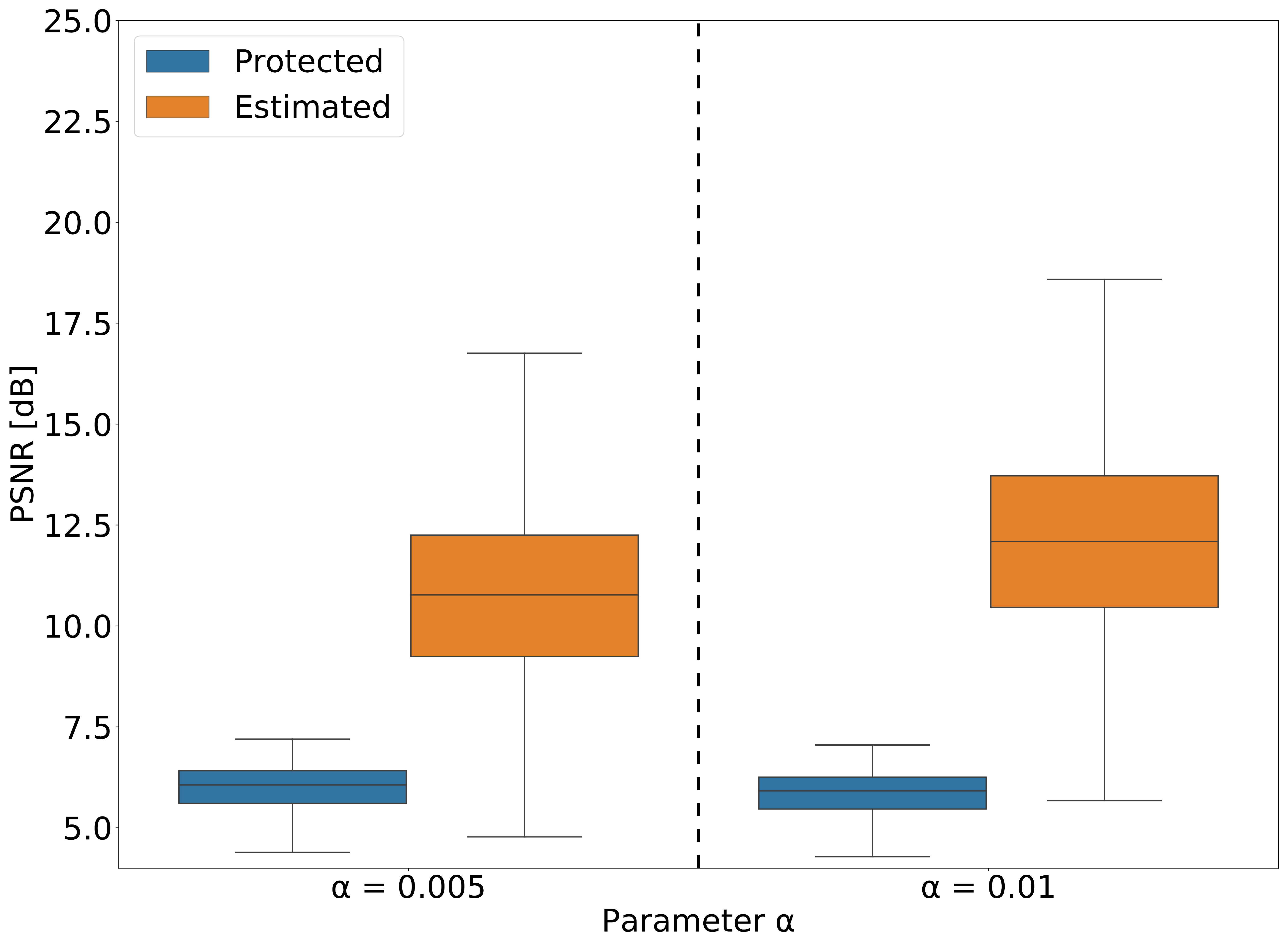}}
  	\caption{PSNR values of estimated images.
  	Boxes span from first to third quartile, referred to as $Q_1$ and $Q_3$, and whiskers show maximum and minimum values in range of [$Q_1 - 1.5(Q_3 - Q_1), Q_3 + 1.5(Q_3 - Q_1)$].
	Band inside box indicates median.
	Outliers are not indicated.}
	\label{fig:restoration}
\end{figure}

Figure \ref{fig:restoration} illustrates PSNR values calculated by using the 10,000 images in the test set of the CIFAR-100 dataset.
The figure shows that the estimated images still had low PSNR values.
In addition, all of the 10,000 estimated images were confirmed to have no visual information on plain images as well as in Fig.\ \ref{fig:enc-stl10}.
From these results, visually protected images are robust against ITN-Attack.

\section{Conclusion}
	In this paper, we proposed a transformation network for generating visually-protected images for privacy-preserving DNNs.
The proposed network enables us not only to protect visual information on plain images but also to maintain high classification accuracy.
Experimental results demonstrated that images generated by the proposed transformation network have almost no visual information.
We also confirmed that the visually-protected images are robust enough against ITN-Attack.

\bibliographystyle{IEEEtran}
\bibliography{strings}

\end{document}